\documentclass{aa}

\usepackage[dvips]{graphicx}
\usepackage{epsfig}
\usepackage{txfonts}
\usepackage{natbib}

\newcommand{\rot}{\vec{\rm rot}\,}
\renewcommand{\div}{{\rm div}\,}

\begin{document}

\title{The magnetron instability in a pulsar's cylindrical
  electrosphere.}

\author{J\'er\^ome P\'etri \inst{1}}

\offprints{J. P\'etri}

\institute{Max-Planck-Institut f\"ur Kernphysik, Saupfercheckweg 1,
  69117 Heidelberg, Germany.}

\date{Received / Accepted}

\titlerunning{Magnetron instability in a pulsar's cylindrical
  electrosphere}

\authorrunning{P\'etri}

\abstract
{The physics of the pulsar magnetosphere near the neutron star surface
  remains poorly constrained by observations. Although about
  2000~pulsars have been discovered to date, little is known about
  their emission mechanism, from radio to high-energy X-ray and
  gamma-rays.  Large vacuum gaps probably exist in the magnetosphere,
  and a non-neutral plasma partially fills the neutron star
  surroundings to form an electrosphere.}
{In several previous works, we showed that the differentially rotating
  equatorial disk in the pulsar's electrosphere is diocotron unstable
  and that it tends to stabilise when relativistic effects are
  included.  However, when approaching the light cylinder, particle
  inertia becomes significant and the electric drift approximation is
  violated.  In this paper, we study the most general instability,
  i.e. by including particle inertia effects, as well as relativistic
  motions.  Electromagnetic perturbations are described in a fully
  self-consistent manner by solving the cold-fluid and Maxwell
  equations.  This general non-neutral plasma instability is called
  the magnetron instability by plasma physicists.}
{We linearise the coupled relativistic cold-fluid and Maxwell
  equations. The non-linear eigenvalue problem for the perturbed
  azimuthal electric field component is solved numerically with
  standard techniques for boundary-value problems like the shooting
  method. The spectrum of the magnetron instability in a non-neutral
  plasma column confined between two cylindrically conducting walls is
  computed for several cylindrical configurations. For a pulsar
  electrosphere, no outer wall exists. In this case, we allow for
  electromagnetic wave emission propagating to infinity.}
{First we checked our algorithm in the low-density limit. We recover
  the results of the relativistic diocotron instability. When the
  self-field induced by the plasma becomes significant, it can first
  increase the growth rate of the magnetron instability. However,
  equilibrium solutions are only possible when the self-electric
  field, measured by the parameter~$s_{\rm e}$ and tending to disrupt
  the plasma configuration, is bounded to an upper limit, $s_{\rm
    e,max}$.  For $s_{\rm e}$ close to but smaller than this value
  $s_{\rm e,max}$, the instability becomes weaker or can be suppressed
  as was the case in the diocotron regime.}
{When approaching the light-cylinder, particle inertia becomes
  significant in the equatorial disk of the electrosphere. Indeed, the
  rest-mass energy density of the plasma becomes comparable to the
  magnetic energy density. The magnetron instability sets in and takes
  over the destabilisation of the stationary flow initiated by the
  diocotron instability close to the neutron star surface. As a
  consequence, the flow in the pulsar inner magnetosphere is highly
  unstable, leading to particle diffusion across the magnetic field
  line. Therefore, an electric current can circulate in the closed
  magnetosphere and feed the wind with charged particles.}
   

\maketitle

\section{INTRODUCTION}

This year, we celebrate the 40th~year of the discovery of the first
pulsar. Nevertheless, the detailed structure of charge distribution
and electric-current circulation in the closed magnetosphere of a
pulsar remains poorly understood. Although it is often assumed that
the plasma fills the space entirely and corotates with the neutron
star, it is on the contrary very likely that it only partly fills it,
leaving large vacuum gaps between plasma-filled regions. The existence
of such gaps in aligned rotators has been very clearly established by
\cite{1985MNRAS.213P..43K, 1985A&A...144...72K}.  Since then, a number
of different numerical approaches to the problem have confirmed their
conclusions, including some work by \cite{1989Ap&SS.158..297R},
\cite{1989Ap&SS.161..187S}, \cite{1993A&A...268..705Z},
\cite{1993A&A...274..319N}, \cite{1994ApJ...431..718T},
\cite{2002ASPC..271...81S}, and ourselves \citep{2002A&A...384..414P}.
This conclusion about the existence of vacuum gaps has been reached
from a self-consistent solution of the Maxwell equations in the case
of the aligned rotator.  Moreover, \cite{2001MNRAS.322..209S} have
shown by numerical modelling that an initially filled magnetosphere
like the Goldreich-Julian model evolves by opening up large gaps and
stabilises to the partially filled and partially void solution found
by \cite{1985MNRAS.213P..43K} and also by \cite{2002A&A...384..414P}.
The status of models of the pulsar magnetospheres, or electrospheres,
has recently been critically reviewed by \cite{2005RMxAC..23...27M}.
A solution with vacuum gaps has the peculiar property that those parts
of the magnetosphere that are separated from the star's surface by a
vacuum region are not corotating and so suffer differential rotation,
an essential ingredient that will lead to non-neutral plasma
instabilities in the closed magnetosphere, a process never addressed
in detail.
  
This raises the question of the stability of such a charged plasma
flow in the pulsar magnetosphere.  The differential rotation in the
equatorial, non-neutral disk induces a non-neutral plasma instability
that is well known to plasma physicists (\citealt{1980PhFl...23.2216O,
  Davidson1990, 1992PhFlB...4.2720O}). Their good confinement
properties (trapped particles can remain on an almost unperturbed
trajectory for thousands of gyro-periods) makes them a valuable tool
for studying plasmas in laboratory, by using for instance Penning
traps.  In the magnetosphere of a pulsar, far from the light cylinder
and close to the neutron star surface, the instability reduces to its
non-relativistic and electrostatic form, the diocotron instability.
The linear development of this instability for a differentially
rotating charged disk was studied by \cite{2002A&A...387..520P}, in
the thin disk limit, and by \cite{2007A&A...469..843P,
  2007A&A...464..135P} in the thick disk limit. It both cases, the
instability proceeds at a growth rate comparable to the star's
rotation rate.  The non-linear development of this instability was
studied by \cite{2003A&A...411..203P}, in the framework of an
infinitely thin disk model. They have shown that the instability
causes a cross-field transport of these charges in the equatorial
disk, evolving into a net out-flowing flux of charges.
\cite{2002ASPC..271...81S} have numerically studied the problem and
concluded that this charge transport tends to fill the gaps with
plasma. The appearance of a cross-field electric current as a result
of the diocotron instability has been observed by
\cite{2002AIPC..606..453P} in laboratory experiments in which charged
particles were continuously injected in the plasma column trapped in a
Malmberg-Penning configuration.
    
A general overview of the equilibrium and stability properties of
non-neutral plasmas in Cartesian and cylindrical geometry can be found
in~\cite{1991RvMP...63..341D}. \cite{1986PhRvA..33.4284T} describe how
to compute fully self-consistent general equilibria configuration for
a cold-fluid plasma in a cylindrical diode. This is useful to
investigate the stability properties in magnetron devices as presented
in \cite{1986PhFl...29.3832D}.

The aim of this work is to extend the previous work by
\cite{2007A&A...469..843P, 2007A&A...464..135P} on the diocotron
instability by including particle inertia effects.  In this paper we
present a numerical analysis of the linear growth of the relativistic
magnetron instability for a non-neutral plasma column.  The paper is
organised as follows.  In Sect.~\ref{sec:Modele}, we describe the
initial setup of the plasma column consisting of an axially symmetric
equilibrium between two conducting walls. We give several equilibrium
profiles useful for the study of the magnetron instability in
different configurations. In Sect.~\ref{sec:AnalLin}, the non-linear
eigenvalue problem satisfied by the perturbed azimuthal electric field
component is derived. The algorithm to solve the eigenvalue problem is
checked against known analytical results in the low-density limit
(diocotron instability), Sect.~\ref{sec:Check}.  Then, applications to
some typical equilibrium configurations are shown in
Sect.~\ref{sec:Results}. First we consider a plasma column with
constant density. Next, we study the effect of the cylindrical
geometry (curvature of the flow) and the transition to the planar
diode limit. Finally, the consequences of the magnetron instability on
the pulsar electrosphere is investigated.  The conclusions and the
possible generalisation are presented in Sect.~\ref{sec:Conclusion}.

\section{THE MODEL}
\label{sec:Modele}

We study the motion of a non-neutral plasma column of infinite axial
extend along the $z$-axis. We adopt cylindrical coordinates denoted
by~$(r,\varphi,z)$ and define the corresponding orthonormal basis
vectors by~$(\vec{e}_{\rm r},\vec{e}_{\rm \varphi},\vec{e}_{\rm z})$.
The geometric configuration is the same as the one described in our
previous works (\citealt{2007A&A...469..843P, 2007A&A...464..135P}).

In this section, we briefly summarise the equilibrium conditions
imposed on the plasma and give some typical examples of equilibrium
configurations for specified velocity, density and electric field
profiles.

We consider a single-species non-neutral plasma consisting of
particles with mass~$m_{\rm e}$ and charge~$q$ trapped between two
cylindrically conducting walls located at the radii $W_1$ and $W_2 >
W_1$. The plasma column itself is confined between the radii $R_1 \ge
W_1$ and $R_2 \le W_2$, with $R_1<R_2$.  This allows us to take into
account vacuum regions between the plasma and the conducting walls.
Because we solve the full set of Maxwell equations, there is also the
possibility for the plasma to radiate electromagnetic waves to
infinity. In order to take this effect into account, we remove the
outer wall if necessary and replace it by outgoing wave boundary
conditions at~$R_2$.

\subsection{Equations of motion}

Let us describe the equation of motion satisfied by the plasma column.
Each particle evolves in the self-consistent electromagnetic field
partly imposed by an external device and partly induced by the plasma
itself. The motion of the plasma column is governed by the cold-fluid
equations, i.e. the conservation of electric charge and the
conservation of momentum, respectively,
\begin{eqnarray}
  \label{eq:Continuite}
  \frac{\partial \rho_{\rm e}}{\partial t} + 
  \div (\rho_{\rm e} \, \vec{v}) & = & 0 \\
  \label{eq:Vit}
  \left( \frac{\partial}{\partial t} + 
    \vec{v} \cdot \frac{\partial}{\partial \vec{r}} \right)
  ( \gamma \, m_{\rm e} \, \vec{v} ) & = & 
  q \, ( \vec{E} + \vec{v} \wedge \vec{B})
\end{eqnarray}
We introduced the usual notation, $\rho_{\rm e}$ for the electric
charge density, $\vec{v}$ for the velocity, $(\vec{E}, \vec{B})$ for
the electromagnetic field. The Lorentz factor of the flow is $\gamma =
1 / \sqrt{1 - \vec{v}^2/c^2}$. Eqs.~(\ref{eq:Continuite}) and
(\ref{eq:Vit}) are supplemented by the full set of Maxwell equations,
\begin{eqnarray}
  \label{eq:MaxFar}
  \rot \vec{E} & = & - \frac{\partial \vec{B}}{\partial t} \\
  \label{eq:MaxAmp}
  \rot \vec{B} & = & \mu_0 \, \vec{j} +
  \varepsilon_0 \, \mu_0 \, \frac{\partial \vec{E}}{\partial t} \\
  \label{eq:MaxGauss}
  \div \vec{E} & = & \frac{\rho_{\rm e}}{\varepsilon_0} \\
  \label{eq:MaxB}
  \div \vec{B} & = & 0
\end{eqnarray}
where $\mu_0$ is the magnetic permeability and $\varepsilon_0$ the
electric permittivity. For a non-neutral plasma, the current density
corresponds to convective motion in the plasma and is therefore
related to the charge density by
\begin{equation}
  \label{eq:DensiteCourant}
  \vec{j} = \rho_{\rm e} \, \vec{v}
\end{equation}
The set of Eqs.~(\ref{eq:Continuite})-(\ref{eq:DensiteCourant})
represents the most general treatment of the cold fluid behavior,
including relativistic and electromagnetic effects as well as inertia
of the particles (the electric drift approximation, used to study the
diocotron instability, is replaced by Eq.~(\ref{eq:Vit})).

\subsection{Equilibrium of the plasma column}

In equilibrium, the particle number density is~$n_{\rm e}(r)$ and the
charge density is~$\rho_{\rm e}(r) = q \, n_{\rm e}(r)$.  Particles
evolve in a cross electric and magnetic field such that the
equilibrium magnetic field is directed along the~$z$-axis whereas the
equilibrium electric field is directed along the $r$-axis. On average,
the stationary motion is only azimuthal. The electric field induced by
the plasma itself is
\begin{eqnarray}
  \label{eq:ErTot}
  \vec{E}_\mathrm{p} & = & E_{\rm r} \, \vec{e}_{\rm r}.
\end{eqnarray}
The magnetic field is made of two parts, an imposed external applied
field, $\vec{B}_0$, assumed to be uniform in the region outside the
plasma column, and a plasma induced field, $\vec{B}_\mathrm{p}$ such
that the total magnetic field is
\begin{equation}
  \label{eq:ETot}
  \vec{B} = \vec{B}_\mathrm{p} + \vec{B}_0 = B_{\rm z} \, \vec{e}_{\rm z} .
\end{equation}
Therefore, for azimuthally symmetric equilibria, the steady-state
Maxwell-Gauss and Maxwell-Amp\`ere equations satisfy
\begin{eqnarray}
  \label{eq:PoissonDiocRelat}
  \frac{1}{r} \, \frac{\partial}{\partial r} ( r \, E_{\rm r}) & = &
  \frac{\rho_{\rm e}}{\varepsilon_0} \\
  \label{eq:AmpereDiocRelat}
  \frac{\partial B_{\rm z}}{\partial r} & = & - \mu_0 \, \rho_{\rm e} \, v_\varphi .
\end{eqnarray}
In the stationary state, $\partial/\partial t = 0$, the balance
between the Lorentz force and the centrifugal force for a fluid
element in Eq.~(\ref{eq:Vit}) is expressed as
\begin{equation}
  \label{eq:vDiocRelat}
  \gamma \, m_{\rm e} \, \frac{v_\varphi^2}{r} + 
  q \, ( E_{\rm r} + v_\varphi \, B_{\rm z} ) = 0 .
\end{equation}
It is convenient to introduce the non-relativistic plasma and cyclotron
frequencies respectively by
\begin{eqnarray}
  \label{eq:Omega_p}
  \Omega_{\rm p}^2 & = & \frac{\rho_{\rm e} \, q}{m_e \, \varepsilon_0} \\
  \label{eq:Omega_c}
  \Omega_{\rm c} & = & \frac{q \, B_{\rm z}}{m_e}
\end{eqnarray}
The corresponding relativistic expressions are respectively
\begin{eqnarray}
  \label{eq:omega_p}
  \omega_{\rm p}^2 & = & \frac{\Omega_{\rm p}^2}{\gamma} \\
  \label{eq:omega_c}
  \omega_{\rm c} & = & \frac{\Omega_{\rm c}}{\gamma}
\end{eqnarray}
where $\gamma = 1 / \sqrt{1 - v_\varphi^2 / c^2}$ corresponds to the
Lorentz factor of the azimuthal flow.  The plasma equilibrium is
governed by two antagonistic effects. On one hand, the electric field
induced by the plasma itself exerts a repelling force on the fluid
element, trying to inflate the plasma. On the other hand, the magnetic
field confines the plasma by incurving the particle trajectories. To
quantify the strength of each effect, focusing magnetic field and
defocusing electric field, it is useful to introduce the self-field
parameter defined by
\begin{equation}
  \label{eq:SelfField}
  s_{\rm e} \equiv \frac{\omega_{\rm p}^2}{\omega_{\rm c}^2}
  = \gamma \, \frac{\Omega_{\rm p}^2}{\Omega_{\rm c}^2}
\end{equation}
Note that this self-field parameter is a local quantity defined on
every point in space as are the plasma and cyclotron frequencies.  It
can therefore depend on the radius~$r$ and is generally not a constant
throughout the plasma column. We should write it $s_{\rm e}(r)$ but in
order to avoid overloading, we do not explicitly show the space
dependence of any quantity as they should all depend on $r$ except
otherwise specified. The diocotron regime, or electric drift
approximation, corresponds to the low-density limit, i.e. to $s_{\rm
  e} \ll 1$ in the whole plasma column. We assume that the electric
field induced by the plasma vanishes at the inner wall, at $r=W_1$,
i.e.
\begin{equation}
  \label{eq:ErCL}
  \vec{E}_\mathrm{p}(W_1) = \vec{0} .  
\end{equation}
Integrating Eq.~(\ref{eq:PoissonDiocRelat}) therefore gives for the
electric field generated by the plasma,
\begin{equation}
  \label{eq:Ep}
  \vec{E}_\mathrm{p}(r) = \frac{1}{\varepsilon_0\,r} \, \int_{W_1}^r 
  \rho_{\rm e}(r') \, r' \, dr' \, \vec{e}_{\rm r}
\end{equation}
For the magnetic field induced by the plasma, we solve
Eq.~(\ref{eq:AmpereDiocRelat}) with the boundary condition $B_{\rm
  z}(R_2) = B_0$. This simply states that the total magnetic field
outside the plasma column has to match the magnetic field imposed by
an external device. Integrating within the plasma column, it is given
by~:
\begin{equation}
  \label{eq:Bzr}
  B_{\rm z}(r) = B_0 - \mu_0 \, \int_{R_2}^r \rho_{\rm e}(r') \,
  v_\varphi(r') \, dr'
\end{equation}

Any equilibrium state is completely determined by the following four
quantities, the total radial electric field, $E_{\rm r}$, the total
axial magnetic field, $B_{\rm z}$, the charge density, $\rho_{\rm e}$,
and the azimuthal speed of the guiding center, $v_\varphi$.
Prescribing one of these profiles, the remaining three are found
self-consistently by solving the set of
Eqs.~(\ref{eq:PoissonDiocRelat}), (\ref{eq:AmpereDiocRelat}) and
(\ref{eq:vDiocRelat}). In the next subsections, we show how to derive
these quantities for some typical examples in which either the
velocity profile, the density profile or the electric field are
imposed.

\subsection{Specified velocity profile}

Let us first assume that the velocity profile~$v_\varphi = r \,
\Omega$ is prescribed, $\Omega$ being the angular velocity at
equilibrium.  This case is well-suited for the study of the pulsar's
electrosphere in which the plasma is in differential rotation and
evolves in a dipolar magnetic field. As already noticed before, the
differential rotation is essential to the presence of the instability.
The other equilibrium quantities, $(E_{\rm r}, B_{\rm z}, \rho_{\rm
  e})$, are easily derived from $\Omega$.  Indeed, inserting
$\rho_{\rm e}$ from Eq.~(\ref{eq:PoissonDiocRelat}) and $E_{\rm r}$
from Eq.~(\ref{eq:vDiocRelat}) into Maxwell-Amp\`ere equation
(\ref{eq:AmpereDiocRelat}), the magnetic field satisfies a first order
linear ordinary differential equation
\begin{eqnarray}
  \label{eq:BzEDO}
  \frac{\partial B_{\rm z}}{\partial r}  & =  & 
  \frac{\gamma^2 \, \beta}{r} \, \left[ B_{\rm z} \, 
  \frac{\partial}{\partial r} ( r \, \beta ) +
  \frac{m_{\rm e}\,c}{q} \, 
  \frac{\partial}{\partial r} \left( \gamma \, \beta^2 \right) \right] .
\end{eqnarray}
The Lorentz factor of the flow is 
\begin{eqnarray}
  \label{eq:LorentzFlow}
  \gamma & = & \frac{1}{\sqrt{ 1 - \beta^2 }} \\
  \beta & = & \frac{r \, \Omega}{c}
\end{eqnarray}
From Maxwell-Amp\`ere equation, Eq.~(\ref{eq:AmpereDiocRelat}), the
charge density is found by
\begin{equation}
  \label{eq:rho2Bz}
  \rho_{\rm e} = - \frac{1}{\mu_0 \, v_\varphi} \, 
  \frac{\partial B_{\rm z}}{\partial r} 
\end{equation}
The electric field is recovered from the force balance equation,
Eq.~(\ref{eq:vDiocRelat}),
\begin{equation}
  \label{eq:Er2Bz}
  E_{\rm r} = - \frac{\gamma \, m_{\rm e} \, v_\varphi^2}{q \, r} -
  v_\varphi \, B_{\rm z}  
\end{equation}

\subsection{Specified density profile}
\label{sec:ProfDens}

Another interesting case corresponds to a specified charge density
profile~$\rho_{\rm e}$. Replacing Eq.~(\ref{eq:Ep}) and (\ref{eq:Bzr})
into the force balance Eq.~(\ref{eq:vDiocRelat}), the rotation profile
is solution of a non-linear Volterra integral equation
\begin{eqnarray}
  \label{eq:Volterra}
  & & \frac{r \, \Omega^2}{\sqrt{1 - r^2 \, \Omega^2 / c^2}} 
  + \frac{1}{r} \, \int_{W_1}^r \Omega_{\rm p}^2(r') \, r' \, dr' \\
  & + & r \, \Omega \, \left[ \Omega_{\rm c0} - 
    \frac{1}{c^2} \, \int_{R_2}^r \Omega_{\rm p}^2(r') \,
    \Omega(r') \, r' \, dr' \right] = 0 \nonumber
\end{eqnarray}
$\Omega_{\rm c0} = q\,B_0/m_{\rm e}$ is the non-relativistic cyclotron
frequency associated to the external magnetic field~$B_0$.  Knowing
$\Omega$, the same procedure as in the previous subsection for a
specified velocity profile is applied, i.e. the electromagnetic field
is calculated according to Eq.~(\ref{eq:BzEDO}) and (\ref{eq:Er2Bz}).

\subsection{Specified electric field}
\label{sec:Electric}

It is also possible to specify the equilibrium radial electric field.
An interesting case is given by
\begin{equation}
  \label{eq:Er0}
  E_{\rm r}(r) = \left \lbrace
    \begin{array}{lcl}
      0  & , & W_1 \le r \le R_1 \\
      - c \, B_0 \, \displaystyle{ \frac{\sinh \alpha \, ( r - R_1 )}
        {\cosh \alpha \, ( R_2 - R1 )} } & , & R_1 \le r \le R_2  \\
      - c\, B_0 \, \displaystyle{ \frac{R_2 \, \tanh \alpha \, ( R_2 - R1 )}{r} }
      & , & R_2 \le r \le W_2
    \end{array}
  \right.
\end{equation}
$\alpha$ is a constant useful to adjust the maximal speed of the
column.  The equilibrium electric field profile, Eq.~(\ref{eq:Er0}),
enables us to investigate the influence of the cylindrical geometry
compared to the planar diode geometry. Indeed, in the limit of small
curvature, $R_2 - R_1 \ll R_1$, the eigenvalue problem in cylindrical
geometry reduces to the planar diode case.  The magnetic field is
solution of an ordinary differential equation
\begin{equation}
  \label{eq:BzEDOEr}
  \frac{\partial B_{\rm z}}{\partial r} = 
  - \frac{\beta}{r \, c} \, \frac{\partial}{\partial r} 
  \left( r \, E_{\rm r} \right)
\end{equation}
The density is easily found from Maxwell-Poisson
Eq.~(\ref{eq:PoissonDiocRelat}) to be
\begin{eqnarray}
  \label{eq:DensElec}
  \rho_{\rm e} & = & n_0 \, q \\
  & = & - \frac{\varepsilon_0 \, B_0 \, \alpha}
  {\cosh \alpha \, ( R_2 - R1 )} \, 
  \left[ \cosh \alpha \, ( r - R1 ) + 
    \frac{\sinh \alpha \, ( r - R1 )}{\alpha \, r} \right] \nonumber
\end{eqnarray}
where $n_0$ is the plasma density within the column.  Finally, the
velocity is given by the centrifugal and Lorentz forces balance,
Eq.~(\ref{eq:vDiocRelat}).

\section{LINEAR ANALYSIS}
\label{sec:AnalLin}

In this section, we derive the eigenvalue problem for the magnetron
instability in the most general case.  We apply the standard linear
perturbation theory.  All scalar perturbations of a physical
quantity~$X$ like electric potential, density, and velocity
components, are expressed by the expansion
\begin{equation}
  \label{eq:Expansion}
  X(r,\varphi,t) = X(r) \, e^{i \, (l\,\varphi - \omega\,t)}
\end{equation}
where $l$ is the azimuthal mode and $\omega$ the eigenfrequency.  A
more general treatment of the perturbation in full 3D allowing for a
stratified vertical structure of the plasma column would need to
introduce a wavenumber $k$, directed along the z-axis, in the phase
term, like $(k\, z + l \, \varphi - \omega \, t)$. Techniques to deal
with this general expansion are discussed in Sect.~3.1 of
\cite{2007A&A...469..843P}.

The decomposition in Eq.~(\ref{eq:Expansion}) only allows for a global
mode to propagate in the plasma. The frequency does not depend on the
radius. Nevertheless, we could imagine to split the column of plasma
in different annular layers~$L_i$, let us say $N$~layers such that
$i\in[[1..N]]$. Each of them possesses its own
eigenfrequency~$\omega_i$. Then, by using matching conditions at the
interface between successive layers, we could solve the eigenvalue
problem and compute the radius dependent growth rates. This would be a
generalisation of the technique employed to match the vacuum solution
to the plasma column as described below.

\subsection{Linearisation of Maxwell equations}

We study the stability of the plasma column around the equilibrium
mentioned in Sect.~\ref{sec:Modele}.  An expansion to first order of
the electromagnetic field around this equilibrium $(\vec{E}^0, B_{\rm
  z}^0)$ leads to
\begin{eqnarray}
  \label{eq:DE}
  \vec{E} & = & \vec{E}^0 + \delta \vec{E} \\
  \label{eq:DBz}
  B_{\rm z} & = & B_{\rm z}^0 + \delta B_{\rm z}   
\end{eqnarray}
and the same for the charge and current densities
\begin{eqnarray}
  \label{eq:Dj}
  \vec{j} & = & \vec{j}^0 + \delta \vec{j} \\
  \label{eq:Drho}
  \rho_{\rm e} & = & \rho_{\rm e}^0 + \delta \rho_{\rm e}   
\end{eqnarray}
Linearising the set of Maxwell equations,
(\ref{eq:MaxFar})-(\ref{eq:MaxB}), we have
\begin{eqnarray}
  \frac{1}{r} \, \frac{\partial}{\partial r} ( r \, \delta \, E_{\rm r}) +
  i \, \frac{l}{r} \, \delta E_\varphi & = & 
  \frac{\delta \rho_{\rm e}}{\varepsilon_0} \\
  \frac{1}{r} \, \frac{\partial}{\partial r} ( r \, \delta \, E_\varphi) -
  i \, \frac{l}{r} \, \delta E_{\rm r} & = & i \, \omega \, \delta B_{\rm z} \\
  i \, \frac{l}{r} \, \delta B_{\rm z} & = & \mu_0 \, \delta j_{\rm r} - 
  i \, \frac{\omega}{c^2} \, \delta E_{\rm r} \\
  - \frac{\partial}{\partial r} \delta B_{\rm z} & = & 
  \mu_0 \, \delta j_\varphi -
  i \, \frac{\omega}{c^2} \, \delta E_\varphi
\end{eqnarray}
The current density perturbation is
\begin{eqnarray}
  \delta j_{\rm r} & = & \rho_{\rm e} \, \delta v_{\rm r} \\
  \delta j_\varphi & = & \delta \rho_{\rm e} \, v_\varphi + 
  \rho_{\rm e} \, \delta v_\varphi
\end{eqnarray}
It is convenient to introduce a potential $\delta E_\varphi = -i \, l
\, \phi / r$ (we emphasise that this function is not the scalar
potential from which the electric field could be derived, but is just
a convenient auxiliary variable) such that the electric and magnetic
field become
\begin{eqnarray}
  \label{eq:deltaEr}
  \delta E_{\rm r} & = & - \kappa \, \left[ 
    \frac{\partial\phi}{\partial r} - i \, \mu_0 \, \omega \,
    \frac{r^2}{l^2} \, \rho_{\rm e} \, \delta v_{\rm r} \right] \\
  \label{eq:deltaBz}
  \delta B_{\rm z} & = & \frac{\omega \, r}{l} \, \kappa \, 
  \left[ \frac{1}{c^2} \, \frac{\partial \phi}{\partial r} - i \,
    \frac{\mu_0}{\omega} \, \rho_{\rm e} \, \delta v_{\rm r} \right] \\
  \kappa & = & \frac{1}{ 1 - ( \omega \, r / l \, c )^2}
\end{eqnarray}
Maxwell-Gauss equation, (\ref{eq:MaxGauss}), is therefore written
\begin{equation}
  \label{eq:MaxGaussLin}
  \frac{1}{r} \, \frac{\partial}{\partial r} 
  \left( r \, \kappa \, \frac{\partial \phi}{\partial r} \right) -
  \frac{l^2}{r^2} \, \phi = - \frac{\delta\rho_{\rm e}}{\varepsilon_0} \, +
  i \, \mu_0 \, \frac{\omega}{r} \, \frac{\partial}{\partial r}
  \left( \frac{r^3}{l^2} \, \kappa \, \rho_{\rm e} \, \delta v_{\rm r} \right)
\end{equation}
In order to find the eigenvalue equation satisfied by the
function~$\phi$, we have to relate the density and velocity
perturbations, $\delta\rho_{\rm e}$, $\delta v_{\rm r}$ and $\delta
v_\varphi$, to~$\phi$. These expressions are derived in the next
paragraph.

\subsection{Linearisation of the fluid equations}

Linearising the conservation of momentum, Eq.~(\ref{eq:Vit}), the
radial and azimuthal perturbations in velocity are related to the
perturbation in electromagnetic field as follows
\begin{eqnarray}
  \label{eq:Vrlin}
  - i \, ( \omega - l \, \Omega ) \, \delta v_{\rm r} - 
  ( \omega_{\rm c} + 2 \, \Omega_{\rm b} ) \, \delta v_\varphi & = & \nonumber \\
  \frac{q}{\gamma \, m_{\rm e}} \, ( \delta E_{\rm r} + 
  r \, \Omega \, \delta B_{\rm z} )
  - i \, ( \omega - l \, \Omega ) \, \gamma^2 \, \delta v_\varphi  & & \nonumber \\
  - \left( \omega_{\rm c} + \frac{1}{\gamma \, r} \, \frac{\partial}{\partial r}
  ( r^2 \, \gamma \, \Omega ) \right) \, \delta v_{\rm r} & = &
  \frac{q}{\gamma \, m_{\rm e}} \, \delta E_\varphi  
\end{eqnarray}
Following \cite{1986PhFl...29.3832D}, we introduced the Coriolis
frequency
\begin{equation}
  \label{eq:Coriolis}
  \Omega_{\rm b} = \frac{1+\gamma^2}{2} \, \Omega
\end{equation}
From Eqs.~(\ref{eq:deltaEr}) and (\ref{eq:deltaBz}), we find
\begin{equation}
  \label{eq:dErdBz}
  \delta E_{\rm r} + r \, \Omega \, \delta B_{\rm z} = 
  - \kappa \, \left[ \left( 1 - \frac{\omega\,r}{l\,c} \, 
      \frac{\Omega\,r}{c} \right) \, \frac{\partial\phi}{\partial r} 
    - i \, \mu_0 \, \rho_{\rm e} \frac{r^2}{l^2} \, ( \omega - l \, \Omega ) 
    \, \delta v_{\rm r} \right]
\end{equation}
From the continuity equation, (\ref{eq:Continuite}), we get
\begin{equation}
  \label{eq:deltarhoe}
  \delta \rho_{\rm e} = \frac{1}{i\, (\omega - l \, \Omega)} \, 
  \left[ \frac{1}{r} \, \frac{\partial}{\partial r} 
    ( r \, \rho_{\rm e} \, \delta v_{\rm r}) +
    i \, \frac{l}{r} \, \rho_{\rm e} \, \delta v_\varphi \right]
\end{equation}
Introducing the function
\begin{eqnarray}
  \label{eq:nub}
  \Delta & = & ( \omega_{\rm c} + 2 \, \Omega_{\rm b} ) \, 
  \left( \omega_{\rm c} + \frac{1}{\gamma \, r} \, \frac{\partial}{\partial r}
    ( r^2 \, \gamma \, \Omega ) \right) \nonumber \\
  & - & ( \omega - l \, \Omega )^2 \, \gamma^2 \, \left( 1 + \kappa \, 
    \frac{\omega_{\rm p}^2 \, r^2}{l^2 \, c^2} \right)
\end{eqnarray}
we solve for the velocity perturbation as
\begin{eqnarray}
  \label{eq:deltavr}
  \delta v_{\rm r} & = & \frac{i \, q}{\gamma \, m_{\rm e} \, \Delta} \,
  \left[ ( \omega - l \, \Omega ) \, \gamma^2 \, \kappa \, 
    \left( 1 - \frac{\omega\,r}{l\,c} \, \frac{\Omega \, r}{c} \right) \,
    \frac{\partial\phi}{\partial r} \right. \nonumber \\ 
  & - & \left. ( \omega_{\rm c} + 2 \, \Omega_{\rm b} ) \, \frac{l}{r} \, \phi \right] \\
  \label{eq:deltavphi}
  \delta v_\varphi & = & \frac{q}{\gamma \, m_{\rm e} \, \Delta} \, 
  \left[ \kappa \, \left( 1 - \frac{\omega\,r}{l\,c} \, 
      \frac{\Omega \, r}{c} \right) \, 
    \left( \omega_{\rm c} + \frac{1}{\gamma \, r} \, \frac{\partial}{\partial r}
      ( r^2 \, \gamma \, \Omega ) \right) \, \frac{\partial\phi}{\partial r}
  \right. \nonumber \\
  & - & \left. ( \omega - l \, \Omega ) \, \left( 1 + \kappa \, 
      \frac{\omega_{\rm p}^2 \, r^2}{l^2 \, c^2} \right) \, \frac{l}{r} \, \phi \right]
\end{eqnarray}
Therefore, the velocity perturbations are expressed in terms of the
function~$\phi$ as well as the density perturbation,
Eq.~(\ref{eq:deltarhoe}).

The eigenvalue problem for $\phi$ is obtained by inserting
Eqs.~(\ref{eq:deltarhoe}), (\ref{eq:deltavr}) and (\ref{eq:deltavphi})
into Eq.~(\ref{eq:MaxGaussLin}).  After some algebraic manipulations,
we arrive at the final expression
\begin{eqnarray}
  \label{eq:EigenValueProblem}
  \frac{1}{r} \, \frac{\partial}{\partial r} 
  \left[ r \, \kappa \, ( 1 + \chi_{\rm r} ) 
    \, \frac{\partial\phi}{\partial r} \right] - 
  \frac{l^2}{r^2} \, ( 1 + \chi_\varphi ) \, \phi & = & \nonumber \\ 
  \frac{l \, \phi}{(\omega - l \, \Omega) \, r} \, \kappa \,
  \left( 1 - \frac{\omega\,r}{l\,c} \, \frac{\Omega \, r}{c} \right) \,
  \frac{\partial}{\partial r} \left( \frac{\omega_{\rm p}^2}{\Delta} 
  \, ( \omega_{\rm c} + 2 \, \Omega_{\rm b} ) \right)
\end{eqnarray}
with the two auxiliary functions
\begin{eqnarray}
  \label{eq:chir}
  \chi_{\rm r} & = & \frac{\gamma^2 \, \kappa \, \omega_{\rm p}^2}{\Delta} \,
  \left( 1 - \frac{\omega\,r}{l\,c} \, \frac{\Omega \, r}{c} \right)^2 \\
  \chi_\varphi & = & \frac{\omega_{\rm p}^2}{\Delta} \,
  \left[ 1 + \kappa \, \frac{\omega_{\rm p}^2 \, r^2}{l^2 \, c^2} + 
    2 \, ( \omega_{\rm c} + 2 \, \Omega_{\rm b} ) \,
    \kappa^2 \, \frac{\omega \, r^2}{l^3 \, c^2} \right]
\end{eqnarray}
The eigenvalue equation~(\ref{eq:EigenValueProblem}) is very general
(\citealt{1986PhFl...29.3832D}).  It describes the motions of small
electromagnetic perturbations around the given equilibrium state,
Eqs.~(\ref{eq:PoissonDiocRelat}) and (\ref{eq:AmpereDiocRelat}), when
inertia is taken into account. Many aspect of the relativistic
magnetron instability can be investigated with this eigenvalue
equation.  In order to solve the eigenvalue problem, boundary
conditions need to be imposed at the plasma/vacuum interface.  They
play a decisive role in the presence or absence of the instability.
How to treat the transition between vacuum and plasma or outgoing wave
solutions is discussed in the next subsection.

\subsection{Boundary conditions}

\label{sec:Boundary}

In laboratory experiments, the plasma is usually confined between
inner and outer conducting walls. However, in pulsar electrospheres,
no such outer device exists to constrain the electric field at the
outer boundary. Radiation from the plasma could propagate into vacuum
to infinity, carrying energy away from the plasma by Poynting flux. To
allow for this electromagnetic wave production by the instabilities
studied in this work, the outer wall is removed. The electromagnetic
field is solved analytically in vacuum and matched to the solution in
the plasma at the plasma/vacuum interface located at $r=R_2$. First we
discuss the situation in which an outer wall exists and then consider
outgoing waves.

\subsubsection{Outer wall}

When vacuum regions exist between the plasma column and the walls,
special care is required at the sharp plasma/vacuum transitions.
Indeed, the right-hand side of Eq.~(\ref{eq:EigenValueProblem}) then
involves Dirac distributions~$\delta(r)$ because the function $f(r) =
\omega_{\rm p}^2 \, ( \omega_{\rm c} + 2 \, \Omega_{\rm b} ) / \Delta$
is discontinuous at the edges of the plasma column, at $R_1$ and
$R_2$.  Moreover, outside the plasma column, this function vanishes
such that $f(r)=0$ for $r<R_1$ and $r>R_2$.

In other words, its derivative has to be computed as
\begin{eqnarray}
  \label{eq:Dirac1}
  \frac{\partial f}{\partial r} = 
  \delta ( r - R_1 ) \, \left. \frac{\partial f}{\partial r} 
  \right|_{r=R_1} -
  \delta ( r - R_2 ) \, \left. \frac{\partial f}{\partial r} 
  \right|_{r=R_2} +
  \left. \frac{\partial f}{\partial r} \right|_{\rm regular}
\end{eqnarray}
where $|_{\rm regular}$ means the regular (or continuous) part of the
derivative, i.e. which does not involve $\delta$ distributions. It
vanishes in the vacuum regions, $r<R_1$ and $r>R_2$. Therefore, the
first order derivative of~$\phi$ is not continuous at these
interfaces. To overcome this difficulty, we decompose the space
between the two walls into three distinct regions:
\begin{itemize}
\item region~I: vacuum space between inner wall and inner boundary of
  the plasma column, with the solution for the electric potential
  denoted by $\phi_\mathrm{I}$, defined for $W_1 \le r \le R_1$~;
\item region~II: the plasma column itself located between $R_1$ and
  $R_2$, solution denoted by $\phi_\mathrm{II}$, defined for $R_1
  \le r \le R_2$~;
\item region~III: vacuum space between the outer boundary of the
  plasma column and the outer wall, solution denoted by
  $\phi_\mathrm{III}$, defined for $R_2 \le r \le W_2$.
\end{itemize}
In regions~I and III, the vacuum solutions satisfy the required
boundary conditions, $\phi_\mathrm{I}(W_1) = 0$ and
$\phi_\mathrm{III}(W_2) = 0$.

The jump in the first order derivative $\partial\phi/\partial r$ at
each plasma/vacuum interface are easily found by multiplying
Eq.~(\ref{eq:EigenValueProblem}) by~$r$ and then integrating around
each discontinuity.  Performing the calculation, we find the jump
at~$R_1$ to be
\begin{eqnarray}
  \label{eq:Jump1}
  \frac{\partial \phi_{\rm II}}{\partial r}(R_1) -
  \frac{\partial \phi_{\rm I }}{\partial r}(R_1)
  & = & \frac{l \, \phi(R_1)}{\omega - l \, \Omega(R_1)} \,
  \left( 1 - \frac{\Omega(R_1) \, R_1}{c} \, \frac{\omega \, R_1}{l \, c} \right) 
  \, \times \nonumber \\
  & \times & \frac{\omega_{\rm p}^2(R_1)}{\Delta(R_1)} \,
  \frac{\omega_{\rm c}(R_1) + 2 \, \Omega_{\rm b}(R_1)}{R_1 \, ( 1 + \chi_{\rm r}(R_1) ) } .
\end{eqnarray}
Similarly, at the outer interface at~$R_2$, we obtain,
\begin{eqnarray}
  \label{eq:Jump2}
  \frac{\partial \phi_{\rm III}}{\partial r}(R_2) -
  \frac{\partial \phi_{\rm II }}{\partial r}(R_2)
  & = & - \frac{l \, \phi(R_2)}{\omega - l \, \Omega(R_2)} \,
  \left( 1 - \frac{\Omega(R_2) \, R_2}{c} \, 
    \frac{\omega \, R_2}{l \, c} \right) \, \times \nonumber \\
  & \times &  \frac{\omega_{\rm p}^2(R_2)}{\Delta(R_2)} \,
  \frac{\omega_{\rm c}(R_2) + 2 \, \Omega_{\rm b}(R_2)}{R_2 \, ( 1 + \chi_{\rm r}(R_2)) } .
\end{eqnarray}

\subsubsection{Outgoing wave solution}

Because of the wall located at~$r=W_2$, the outer boundary condition
$\phi_\mathrm{III}(W_2) = 0$ enforces $E_\varphi(W_2)=0$.  It
therefore prevents waves escaping from the system due to the vanishing
outgoing Poynting flux, $E_\varphi\,B_z/\mu_0=0$. In pulsar
magnetospheres, no such wall exists. Thus, in order to let the system
produce outgoing electromagnetic waves, we remove the outer wall in
this case and solve the vacuum wave equation for~$\phi$, which then
reads
\begin{equation}
  \label{eq:OndePhiVide}
  \frac{1}{r} \, \frac{\partial}{\partial r} 
  \left[ r \, \kappa(r,\omega) \, \frac{\partial \phi}{\partial r} \right] -
  \frac{l^2}{r^2} \, \phi = 0.
\end{equation}
This equation can also be derived directly from the vector wave
equation
\begin{equation}
  \label{eq:VectorWave}
  \Delta \vec{E} - \frac{1}{c^2} \, 
  \frac{\partial^2 \vec{E}}{\partial t^2} = \vec{0}
\end{equation}
projected along the $e_\varphi$ axis.  To find the right outgoing wave
boundary conditions, it is therefore necessary to solve the
vector-wave equation in cylindrical coordinates using vector
cylindrical harmonics as described for instance in \cite{Stratton1941}
and \cite{1953mtp..book.....M}. The solutions for the function~$\phi$
to be an outgoing wave in a vacuum outside the plasma column, which
vanishes at infinity, is given by (region~III with $W_2 = +\infty$)
\begin{equation}
  \label{eq:PhiOutgoing}
  \phi_{\rm III} = K \, r \,\frac{\partial}{\partial r} 
  H_l \left( \frac{\omega\,r}{c} \right) =
  K \, \frac{\omega \, r}{c} \, H_l' \left( \frac{\omega\,r}{c} \right),
\end{equation}
where the cylindrical outgoing wave function is given by $H_l$
\citep{Stratton1941}, the Hankel function of the first kind and of
order~$l$ related to the Bessel functions by $H_l(x) = J_l(x) + i \,
Y_l(x)$, \citep{1965hmfw.book.....A}. The prime~$'$ means derivative
of the function evaluated at the point given in parentheses, and $K$
is a constant to be determined from the boundary condition at $R_2$.
Eliminating the constant~$K$, we conclude that the boundary condition
to impose on $\phi$ is
\begin{eqnarray}
  \label{eq:CLPhiR2}
  \left[ H_l'\left( \frac{\omega\,R_2}{c} \right) + 
    \frac{\omega\,R_2}{c} \, H_l''\left( \frac{\omega\,R_2}{c} \right) \right] 
  \, \phi_{\rm III}(R_2) & - & \nonumber \\
  R_2 \, H_l'\left( \frac{\omega\,R_2}{c} \right) \, 
  \frac{\partial\phi_{\rm III}}{\partial r}(R_2) & = & 0.
\end{eqnarray}
The boundary conditions {\it expressed in region~II for~$\phi_{\rm
    II}$} are found by replacing $\phi_{\rm III}'(R_2)$ from
Eq.~(\ref{eq:Jump2}). Since $\phi$ is continuous, $\phi_{\rm III}(R_2)
= \phi_{\rm II}(R_2) = \phi_{\rm}(R_2)$. We find
\begin{eqnarray}
  \label{eq:Jump3}
  & & \left[ H_l'\left( \frac{\omega\,R_2}{c} \right) + 
    \frac{\omega\,R_2}{c} \, H_l''\left( \frac{\omega\,R_2}{c} \right) \right] 
  \, \phi(R_2) - \nonumber \\
  & & R_2 \, H_l'\left( \frac{\omega\,R_2}{c} \right) \,
  \left[ \frac{\partial\phi_{\rm II }}{\partial r}(R_2) -  
    \frac{l \, \phi(R_2)}{\omega - l \, \Omega(R_2) } \,
     \right. \times \nonumber \\
  & & \times \left. \left( 1 - \frac{\Omega(R_2) \, R_2}{c} \, 
      \frac{\omega \, R_2}{l \, c} \right) \, \frac{\omega_{\rm p}^2(R_2)}{\Delta(R_2)} \,
  \frac{\omega_{\rm c}(R_2) + 2 \, \Omega_{\rm b}(R_2)}{R_2 \, ( 1 + \chi_{\rm r}(R_2)) } \right]
  = 0 .
\end{eqnarray}

\subsection{Set of annular layers}
  
Because the instability is related to strong velocity gradients, it is
expected that the instability will only exist in the vicinity of this
shear. In this paragraph, we briefly describe how to compute the
radius dependent growth rate without going into details.
  
As was already done for the plasma/vacuum interface, the plasma column
is divided into a set of $N$~annular layers~$L_i$, each of them having
their own eigenfrequency~$\omega_i$ and a radial extension from
$R_1^i$ to $R_2^i > R_1^i$. Because no surface charge accumulates on
the interfaces between successive layers, the matching conditions
require the unknown function~$\phi$ to be continuous as well as its
first derivative when crossing the interface. Moreover, to avoid
shearing in the perturbation, we should impose ${\rm Re}(\omega_i)$ to
have the same constant value in all layers and just look at the
variation of the growth rate, ${\rm Re}(\omega_i)$ is related to the
pattern speed of the perturbation.  As a result, we would expect that
the growth rate depends on radius and is maximal where the shear is
largest. The column will then split into weakly (almost no shear in
the flow) and strongly (large velocity gradients) unstable regions.

\subsection{Algorithm}
\label{sec:Algo}

The eigenvalue problem, Eq.~(\ref{eq:EigenValueProblem}), is solved by
standard numerical techniques. We have implemented a shooting method
as described in \cite{2007A&A...469..843P} by replacing the eigenvalue
Eq.~(54) there, by the new ordinary differential equation,
Eq.~(\ref{eq:EigenValueProblem}) and using the jump conditions
Eq.~(\ref{eq:Jump1}) and~(\ref{eq:Jump2}), when inner and outer walls
are present.

For the pulsar electrosphere, the situation is very similar, except
that no calculation is performed in region~III. The boundary condition
for outgoing waves is applied at $R_2$, see Eq.~(\ref{eq:Jump3}).
Actually for pulsars, we compare both boundary conditions.

\section{Algorithm check}
\label{sec:Check}

In order to check our algorithm in different configurations, we
compute the eigenvalues for both, a low-density relativistic plasma
column and in the limiting case of a relativistic planar diode
geometry.  For some special density profiles, the exact analytical
dispersion relations are known and used as a starting point or for
comparison with the numerical results obtained by our algorithm.

\subsection{Low-density plasma column}
\label{sec:ColNRel}

In cylindrical geometry, an exact analytical solution of the
dispersion relation can be found in the low-density and
non-relativistic limit, the diocotron instability
(\citealt{Davidson1990}).  We use these results to check our algorithm
in cylindrical coordinates, as was already done in
\cite{2007A&A...469..843P}.

For completeness, we briefly recall the main characteristics of the
configuration. The magnetic field is constant and uniform outside the
plasma column, namely $B_{\rm z} = B_0$, for $r>R_2$. We solve
self-consistently the Maxwell equations in the space between the inner
and the outer wall, $R_1 < r < R_2$.  The particle number density and
charge density are constant in the whole plasma column such that
\begin{equation}
  \label{eq:RhoDiocRect}
  \rho_{\rm e}(r) = \left \lbrace
    \begin{array}{lcl}
      0   & , & W_1 \le r \le R_1 \\
      \rho_0 = n_0 \, q = {\rm const} & , & R_1 \le r \le R_2 \\
      0   & , & R_2 \le r \le W_2
    \end{array}
  \right.
\end{equation}
We first check our fully relativistic and electromagnetic code, which
includes particle inertia, in the non-relativistic and low-density
limit. By non-relativistic, we mean a maximum speed at the outer edge
of the plasma column of $\approx 10^{-3}$ and a density such that the
self-field parameter is weak, $s_{\rm e} = 10^{-6}$. When numerical
values of the self-field parameter are given, we mean the value it
takes at the outer edge of the plasma column. Thus $s_{\rm e}$ should
be understood as $s_{\rm e}(r=R_2)$. In this limit, the electric drift
approximation is excellent and the non-relativistic diocotron regime
applies. The eigenvalues are therefore given by Eq.~(68) of
\cite{2007A&A...469..843P}.

We computed the growth rates for several geometrical configurations of
the plasma column, varying $R_{1,2}$ and $W_{1,2}$. Some typical
examples are presented in Table~\ref{tab:DiocCylNRel}. To summarise,
the relative error in the real and imaginary part of the eigenvalues
compared to the exact analytical solution is
\begin{eqnarray}
  \label{eq:Erreur}
  \varepsilon_{\rm Re} & = & 
  \left| \frac{{\rm Re} \, (\omega) - {\rm Re} \, (\omega_{\rm exact})}
    {{\rm Re} \, (\omega_{\rm exact})} \right| \lesssim 10^{-7} \\
  \varepsilon_{\rm Im} & = & 
  \left| \frac{{\rm Im} \, (\omega) - {\rm Im} \, (\omega_{\rm exact})}
    {{\rm Im} \, (\omega_{\rm exact})} \right| \lesssim 10^{-7}
\end{eqnarray}
The precision is excellent, reaching 7~digits at least. Our algorithm
computes quickly and accurately the eigenvalues in cylindrical
geometry with vacuum gaps between the plasma column and the walls.
The eigenvalues obtained in this example are good initial guesses to
study the relativistic problem in the low speed limit.
\begin{table*}[htbp]
  \centering
  \begin{tabular}{cccccc}
    \hline
    Mode $l$ & $d_1$ & $d_2$ & $\omega_{\rm num}$ & 
    $\varepsilon_{\rm Re}$ & $\varepsilon_{\rm Im}$ \\
    \hline
    2 & 0.4  & 0.5 & 1.8865e-03 + 3.5878e-04 \, i & 6.5774e-08 & 4.1221e-08 \\
    3 & 0.4  & 0.5 & 2.7284e-03 + 1.1336e-03 \, i & 7.8227e-08 & 1.3019e-08 \\
    4 & 0.4  & 0.5 & 3.6081e-03 + 1.4944e-03 \, i & 8.6291e-08 & 1.2482e-08 \\
    5 & 0.45 & 0.5 & 2.3766e-03 + 1.3515e-03 \, i & 1.9263e-08 & 3.1651e-09 \\
    7 & 0.45 & 0.5 & 3.3251e-03 + 1.7069e-03 \, i & 2.2522e-08 & 4.2726e-09 \\
    \hline
  \end{tabular}
  \caption{Numerical eigenvalues~$\omega_{\rm num}$ and relative errors
    for the low-density and non-relativistic plasma column,
    for different modes~$l$ and 
    different aspect ratios, $d_1 = R_1 / W_2$, and $d_2 = R_2 / W_2$.}
  \label{tab:DiocCylNRel}  
\end{table*}

As in \cite{2007A&A...469..843P}, the influence of the relativistic
and electromagnetic effects are investigated by slowly increasing the
maximal speed at the outer edge of the plasma column, $\beta_2 = R_2
\, \Omega(R_2) / c$. We also allow for outgoing electromagnetic wave
radiation at the outer edge of the column.

\begin{figure*}[htbp]
  \centering
  \begin{tabular}{cc}
    \includegraphics[scale=0.8]{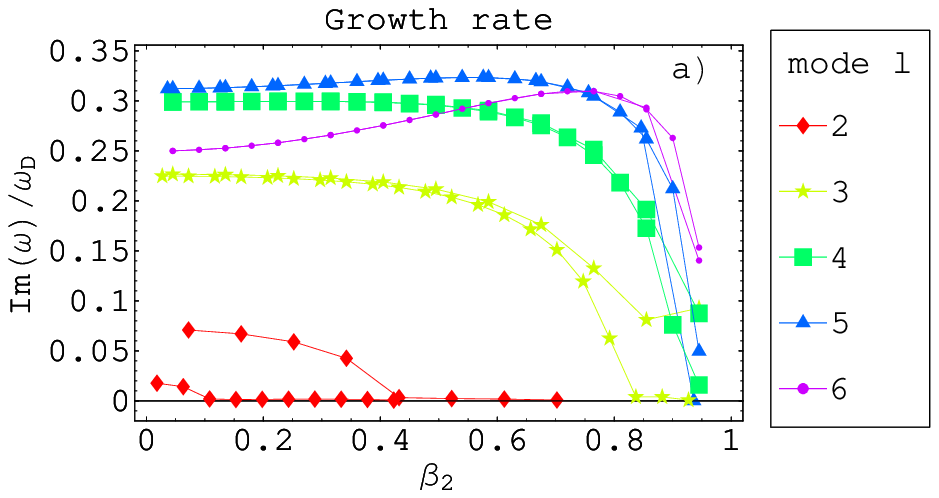} 
    \includegraphics[scale=0.8]{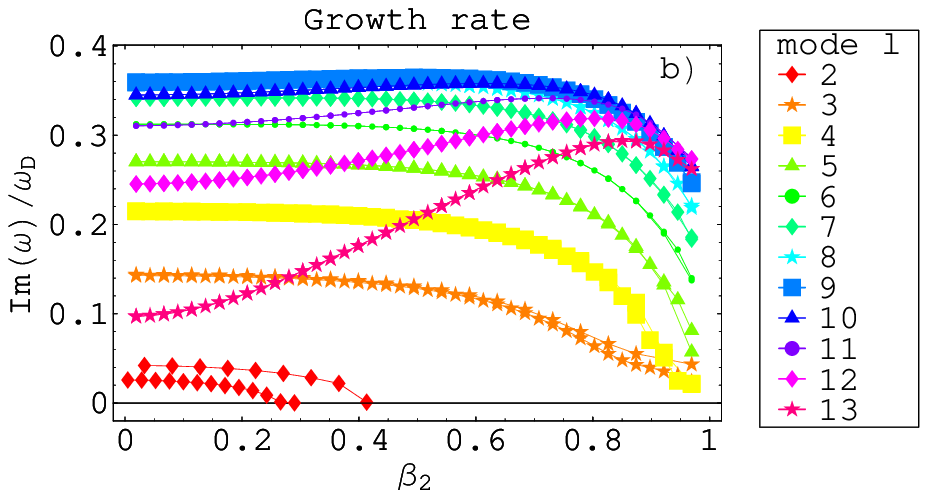}
  \end{tabular}
  \caption{Stabilisation of the magnetron instability in the low-density limit. 
    We retrieve the results of the diocotron instability. The plasma
    density is constant in the column, $\rho_0 = {\rm cst}$.  The
    geometric aspect ratios are, for Fig.(a), $w=0.1$, $d_1=0.4$ and
    $d_2=0.5$ and, for Fig.(b), $w=0.1$, $d_1=0.45$ and $d_2=0.5$. The
    growth rates are normalised to the diocotron frequency,
    $\omega_{\rm D}$. The curves corresponding to the outgoing wave
    boundary conditions contain twice as much symbols as the
    conducting wall boundary conditions in order to distinguish them.}
  \label{fig:ColRel1}
\end{figure*}

Two cases are presented in Fig.~\ref{fig:ColRel1}. The first one,
Fig.~\ref{fig:ColRel1}a, has $w = W_1 / W_2 = 0.1$, $d_1 = R_1 / W_2 =
0.4$ and $d_2 = R_2 / W_2 = 0.5$ whereas the second one,
Fig.~\ref{fig:ColRel1}b, has $w = 0.1$, $d_1 = 0.45$ and $d_2 = 0.5$.
For non-relativistic speeds, $\beta_2 \ll 1$, the eigenvalues of
Sect.~\ref{sec:ColNRel} are recovered. The thinner the plasma layer,
the larger the number of unstable modes, respectively 5 and 12
unstable modes. In both cases, the growth rate starts to be altered
whenever $\beta_2 \gtrsim 0.1$.  In any case, for very high speeds,
$\beta_2 \approx 1$, all the modes stabilise because the growth rate
vanishes. 

Removing the outer wall, and replacing it by outgoing wave boundary
conditions does not significantly affect the instability. Except for
the mode $l=2$, the changes become perceptible only for relativistic
speeds, $\beta_2 \gtrsim 0.7$.

Note that these results slightly differ from those presented in
\cite{2007A&A...469..843P} because in the former case the density is
kept constant while in the latter case the diocotron frequency,
$\omega_{\rm D}$, is kept constant, both being related by the Lorentz
factor of the flow.

A simplistic approach to understand the stabilisation process is as
follows. In the low-density limit, the magnetron instability reduces
to the diocotron regime. Assuming a constant charge density within the
plasma, unstable oscillations are generated when the surface waves at
both edges of the plasma column interact in a constructive way, i.e.
that oscillation have to synchronise.  For high velocities within the
fluid, the relative phase speed between the two waves increase, mutual
synchronisation becomes difficult to reach, reducing the growth rate
of the instability \citep{1966JAP....37..602K}.
  
The qualitative physical nature of the diocotron/magnetron instability
can be expressed via the more familiar Kelvin-Helmholtz or two-stream
instability \citep{1950PPSB...63..409M, 1959JAP....30.1784T,
  1966JAP....37.3203B}. For instance, the two-stream instability is
investigated in \cite{1994DelcroixT1}. It is found (in Cartesian
geometry) that while in the long wavelength limit, the growth rate is
proportional to the velocity difference between the two beams, they
are reduced for sufficiently large velocity shear until the
instability disappears. This stabilisation mechanism applies also to
the diocotron/magnetron instability and can already occur at modest
speeds, much less than the speed of light~$c$.

\subsection{Influence of curvature}

The influence of the curvature is also studied by taking the limit of
the planar diode geometry. The curvature of the plasma column is then
increased to investigate the evolution of the growth rates.

The effective aspect ratio of the plasma layer is conveniently
described by the parameter
\begin{equation}
  \label{eq:AspectA}
  A = \frac{R_1}{R_2 - R_1}
\end{equation}
Using the equilibrium electric field profile indicated in
Sect.~\ref{sec:Electric}, in the limit of small curvature corresponding
to large aspect ratio, $A\to+\infty$, the eigenvalue problem and
equilibrium configuration is described by the relativistic planar
diode.

\begin{figure*}[htbp]
  \centering
  \begin{tabular}{cc}
    \includegraphics[scale=0.8]{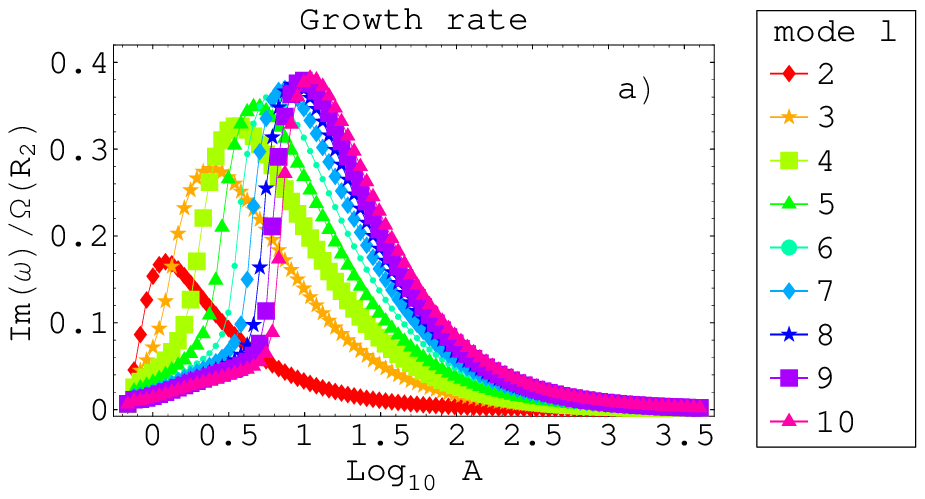} &
    \includegraphics[scale=0.8]{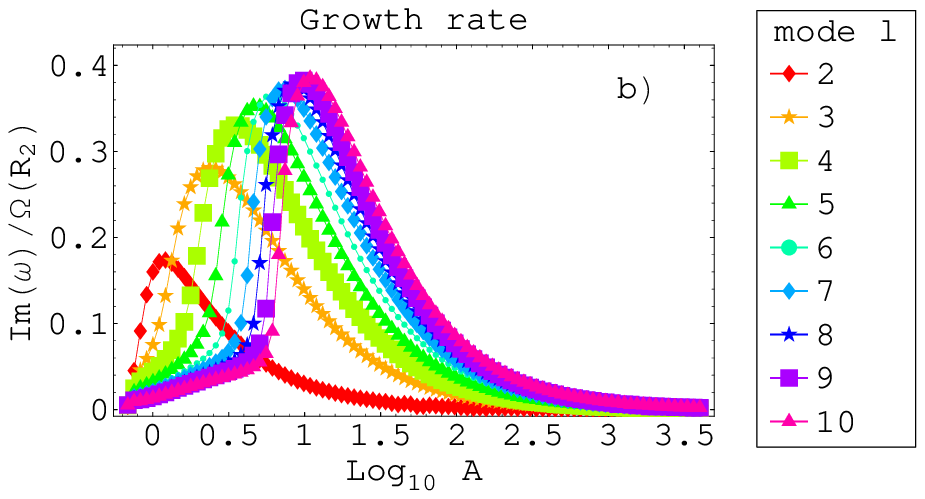}
  \end{tabular}
  \caption{Effect of the cylindrical geometry on the growth rate of the
    instability. The growth rate are normalised to the rotation at the
    outer edge of the plasma column~$\Omega(R_2)$ and plotted versus
    the aspect ratio~$A$, in logarithmic scale. In Fig.(a), the speed
    at the outer edge of the column is $\beta_2 = 10^{-3}$ whereas in
    Fig.(b), it is $\beta_2 \approx 0.2$.}
  \label{fig:Courbure}
\end{figure*}

We show the evolution of the growth rate in the non-relativistic limit
$\beta_2 = 10^{-3}$, Fig.~\ref{fig:Courbure}a, and in the relativistic
case $\beta_2 \approx 0.2$, Fig.~\ref{fig:Courbure}b.  The aspect
ratio has a drastic influence on the growth rate. For large values
of~$A \gg 1$, all unstable modes are stabilised, in both
non-relativistic and relativistic flows.  These results agree with
those found in \cite{2007A&A...469..843P}. Note that the drift speed
has only a negligible effect on the growth rate compared to the aspect
ratio.

\subsection{Electrosphere}

Finally, we checked the results obtained for the electrosphere, with
outer wall or outgoing waves. The rotation profile is chosen to mimic
the rotation curve obtained in the 3D electrosphere.  To study the
influence of the relativistic effects, we take the same profiles as
those given in \cite{2007A&A...469..843P, 2007A&A...464..135P}.  We
remind that different analytical expressions for the radial dependence
of $\Omega$ are chosen by mainly varying the gradient in differential
shear as follows
\begin{equation}
  \label{eq:ProfilVit}
  \Omega(r) = \Omega_* \, (2 + \tanh[  \alpha \, ( r - r_0 ) ] \, e^{-\beta\,r^4})
\end{equation}
$\Omega_*$ is the neutron star spin and $r$ is normalised to the
neutron star radius, $R_*$. The values used are listed in
Table~\ref{tab:Vitesse}.
\begin{table}[htbp]
  \centering
  \begin{tabular}{cccc}
    \hline
    $\Omega$ & $\alpha$ & $\beta$ & $r_0$ \\
    \hline
    $\Omega_2$ & 1.0 & $5\times10^{-5}$ & 6.0 \\
    $\Omega_3$ & 0.3 & $5\times10^{-5}$ & 10.0 \\
    \hline
  \end{tabular}
  \caption{Parameters for the rotation profiles used
    to mimics the azimuthal velocity of the plasma 
    in the electrospheric disk.}
  \label{tab:Vitesse}
\end{table}
\begin{figure*}[htbp]
  \centering
  \begin{tabular}{cc}
  \includegraphics[scale=0.8]{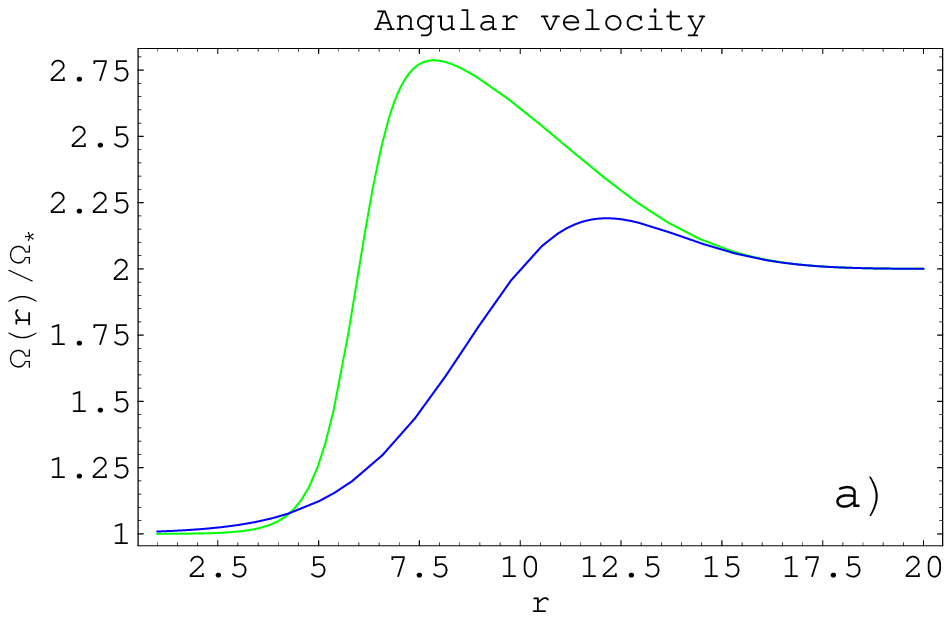} &
  \includegraphics[scale=0.8]{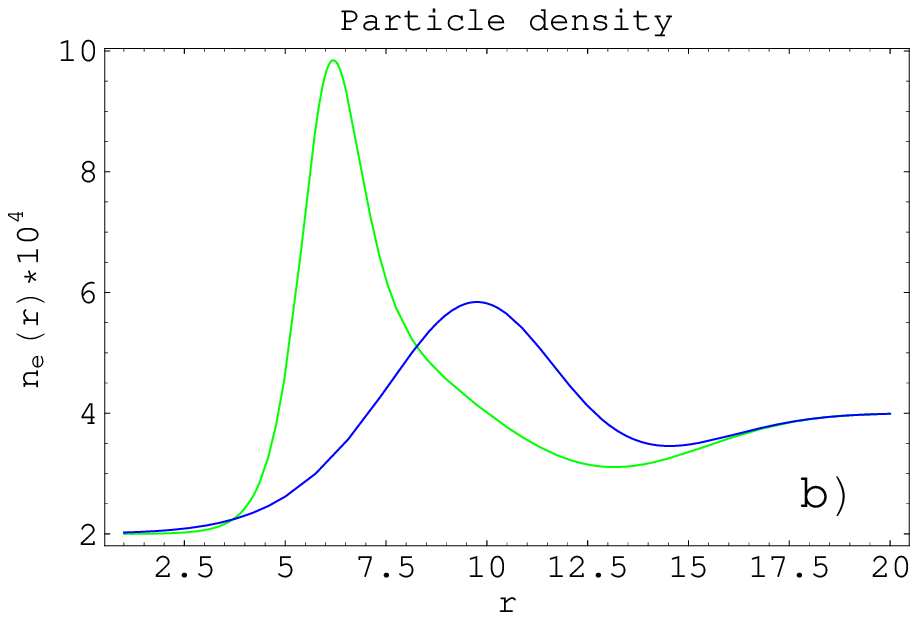}
  \end{tabular}
  \caption{Two choices of differential rotation curves
    in the plasma column for the cylindrical pulsar electrosphere,
    $\Omega_2$ with strong shear (in green) and $\Omega_3$ with weak
    shear (in blue), Fig.~(a) and the corresponding charge density,
    Fig.(b).}
  \label{fig:OmegaElec}
\end{figure*}
The angular velocity, and the corresponding particle density number,
are shown respectively in Fig.~\ref{fig:OmegaElec}a and
\ref{fig:OmegaElec}b.  In both cases, $\Omega$ starts from corotation
with the star $\Omega = \Omega_*$ followed by a sharp increase around
$r=6$ for $\Omega_2$ and a less pronounced gradient around $r=10$ for
$\Omega_3$.  Finally the rotation rate asymptotes twice the neutron
star rotation speed for large radii.

\begin{figure*}[htbp]
  \centering 
  \begin{tabular}{cc}
    \includegraphics[scale=0.8]{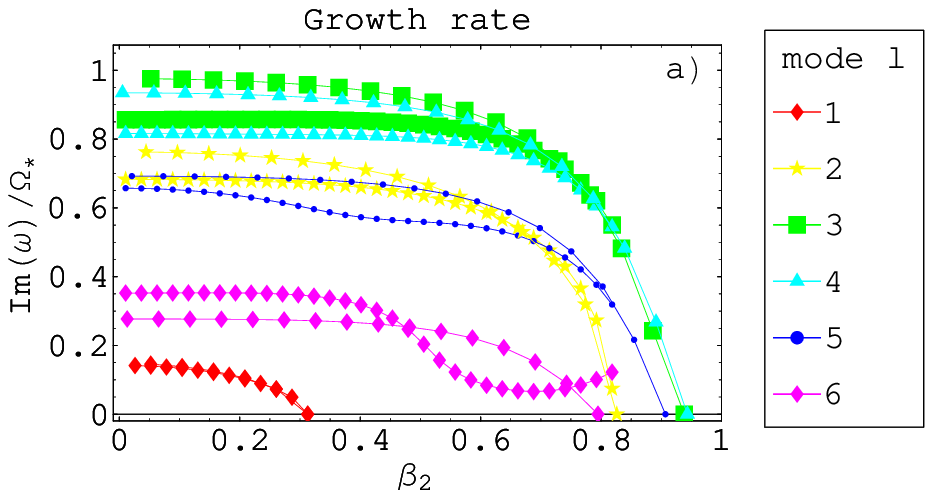} & 
    \includegraphics[scale=0.8]{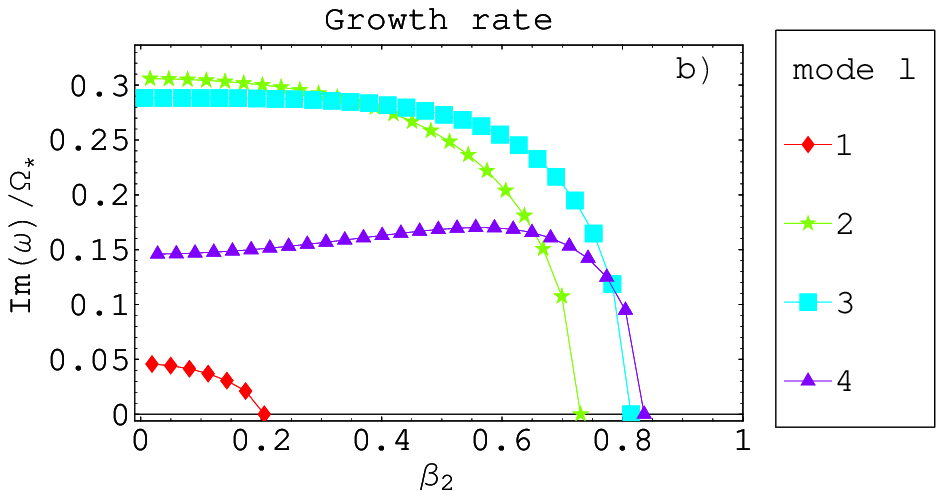}
  \end{tabular}
  \caption{Evolution of the growth rates,~${\rm Im}(\omega)$, 
    normalised to the spin of the neutron star~$\Omega_*$, for
    increasing maximal speed of the column, $\beta_2$. For the profile
    $\Omega_2$, in Fig.(a), outer conducting wall are compared with
    outgoing wave boundary conditions. For the latter, the curves
    contain twice as much symbols as for the former boundary
    conditions.  Note that for the mode $l=1$, the curves overlap and
    cannot be distinguished. For the profile $\Omega_3$, in Fig.(b),
    only the wall boundary conditions are plotted.}
  \label{fig:Elec}
\end{figure*}

The computed growth rates, normalised to the speed of the neutron
star, are shown in Fig.~\ref{fig:Elec}, for the profile $\Omega_2$,
Fig.~\ref{fig:Elec}a, and for $\Omega_3$, Fig.~\ref{fig:Elec}b.

\section{RESULTS}
\label{sec:Results}

We demonstrated that our numerical algorithm gives accurate results in
the non-relativistic cylindrical geometry for which we know an
analytical expression of the eigenfrequencies. Moreover, the
transition to the relativistic regime gives the same results as those
shown in \cite{2007A&A...469..843P}.  In this section, we compute the
eigenspectra of the relativistic magnetron instability in cylindrical
coordinates, for various equilibrium density profiles, electric field,
and velocity profiles.  Application to pulsar's electrosphere is also
discussed.

\subsection{Influence of the self-field}
\label{subsec:Lab}

We study the influence of the self-electromagnetic field on the
instability. When the self-field is negligible, i.e. in the
low-density limit, $s_{\rm e} \ll 1$, the instability is well
approximated by the diocotron regime, which means in the electric
drift approximation. However, when the density of the plasma
increases, $s_{\rm e} \approx 1$, it induces a strong electric field
which would lead to superluminal motion in the drift approximation.
Therefore, particle inertia comes into play and imposes a motion which
departs significantly from the electric drift and modifies the
behavior of the instability.

We considered the plasma in cylindrical geometry, confined by some
external experimental electromagnetic device between the inner and
outer wall. The external applied magnetic field and the density
profile are specified as initial data. Thus, we deduce the velocity
profile by solving the non-linear Volterra integral equation,
Eq.~(\ref{eq:Volterra}), as explained in Sect.~\ref{sec:ProfDens}.

We study the influence of particle inertia in both, the
non-relativistic and relativistic regime. As a starting point, we take
the diocotron regime, $s_{\rm e} \ll 1$, and slowly increase the
influence of the self-field parameter, $s_{\rm e}$.

\begin{figure*}[htbp]
  \centering
  \begin{tabular}{cc}
  \includegraphics[scale=0.8]{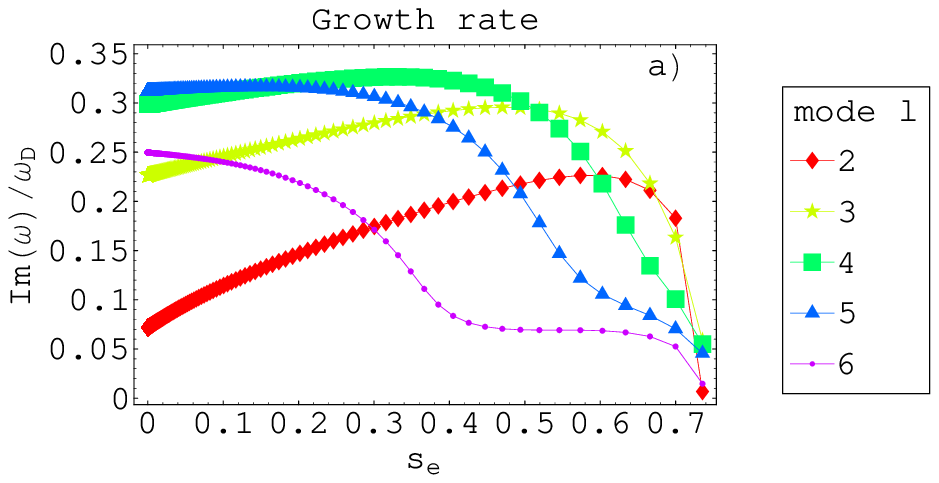} &
  \includegraphics[scale=0.8]{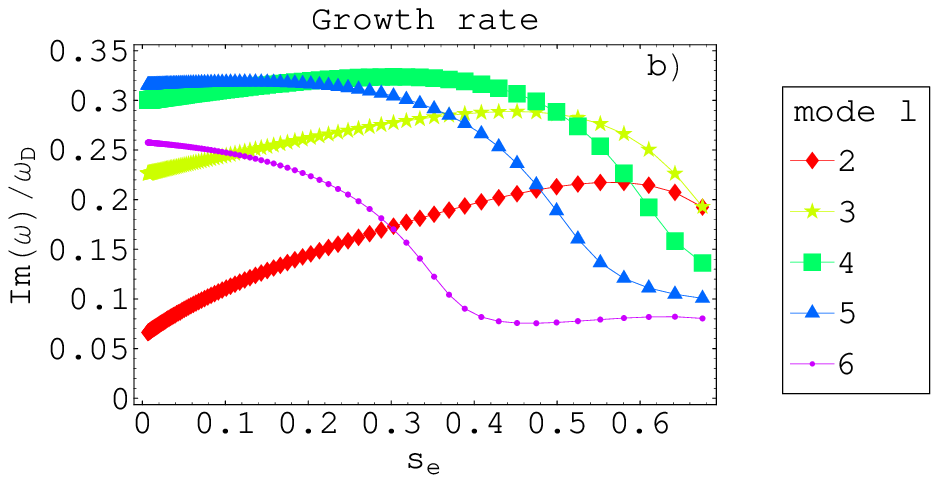}
  \end{tabular}
  \caption{Influence of the self-field parameter, $s_{\rm e}$, 
    on the magnetron instability, for a non-relativistic flow
    ($\beta_2 \approx 10^{-3}$) in Fig.(a) and a relativistic flow
    ($\beta_2 \approx 0.19$) in Fig.(b).  The geometric aspect ratios
    are, $w=0.1$, $d_1=0.4$, and $d_2=0.5$, the same as those in
    Fig.~\ref{fig:ColRel1}a.}
  \label{fig:Auto_Champ}
\end{figure*}

Examples of growth rates are shown in Fig.~\ref{fig:Auto_Champ}.  A
non-relativistic flow ($\beta_2 \approx 10^{-3}$) is shown in
Fig.~\ref{fig:Auto_Champ}a whereas a mildly relativistic flow
($\beta_2 \approx 0.19$) is shown in Fig.~\ref{fig:Auto_Champ}b.  In
both cases, increasing the self-field stabilises the magnetron regime.
Close to the maximum value of the self-field parameter, the Brillouin
zone above which no equilibrium configuration exists because the
defocusing electric field is to strong compared to the magnetic
focusing field, the instability has almost completely disappeared.

\begin{figure*}[htbp]
  \centering
  \begin{tabular}{cc}
  \includegraphics[scale=0.8]{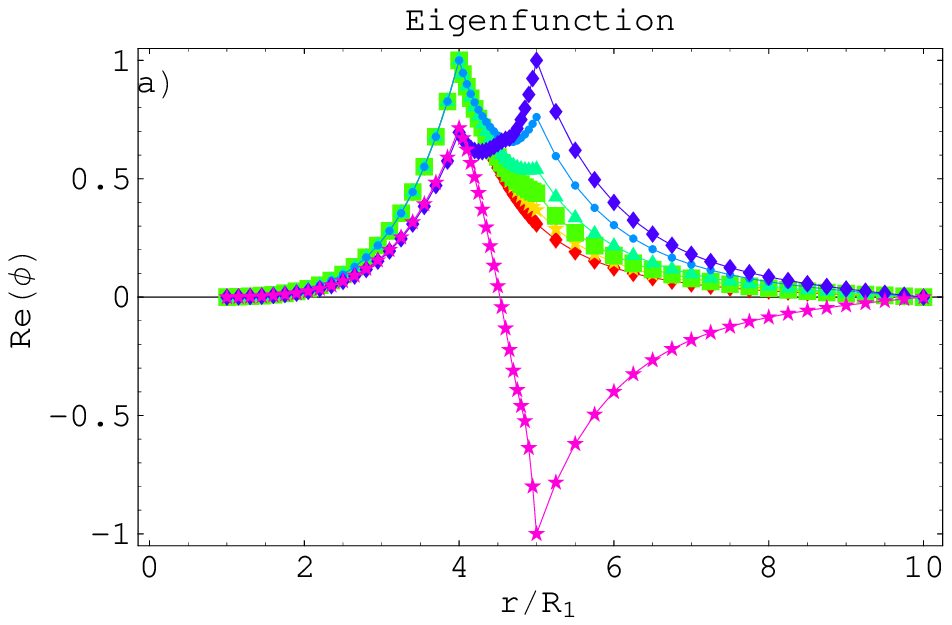} &
  \includegraphics[scale=0.8]{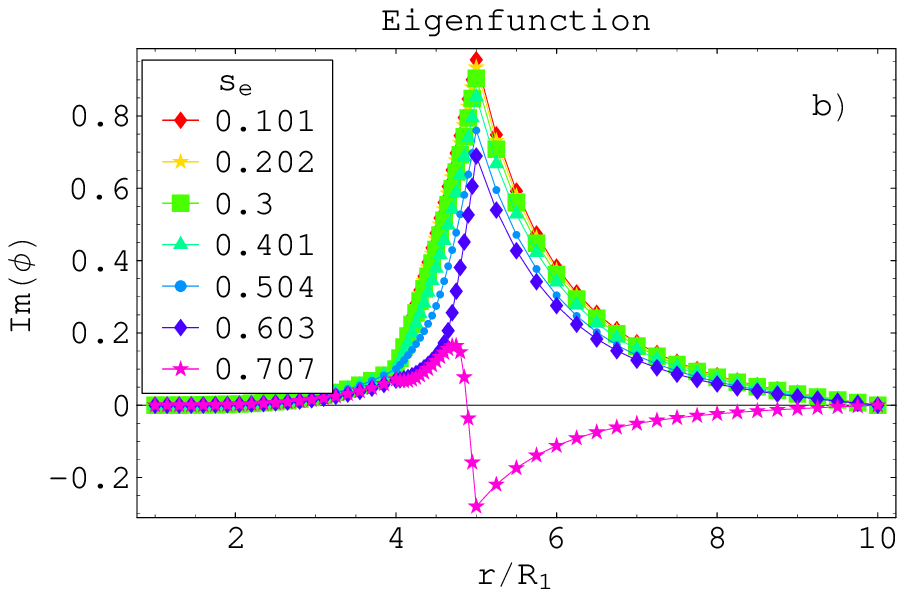}
  \end{tabular}
  \caption{Real and imaginary part of the eigenfunction, 
    respectively Fig.~a and b, for the mode $l=5$ in
    Fig.~\ref{fig:Auto_Champ}a. The real part of the eigenfunction is
    normalised such that the maximum in absolute value is unity. The
    value of the self-field parameter~$s_{\rm e}$ for each
    eigenfunction is shown in legend.}
  \label{fig:Auto_Champ_Vec_Prop}
\end{figure*}

For completeness, the evolution of the shape of the eigenfunctions of
the mode $l=5$ corresponding to the eigenvalues in
Fig.~\ref{fig:Auto_Champ}a is shown in
Fig.~\ref{fig:Auto_Champ_Vec_Prop}. When the stabilisation process
becomes important, the real and imaginary parts of the eigenfunction
possess significant negative values compared to the low self-field
parameter case.

\subsection{Electrosphere}
\label{subsec:Pulsar}

The electrospheric non-neutral plasma is confined by the rotating
magnetised neutron star. The most important feature is the velocity
profile in the plasma column. For simplicity, here, we assume that no
vacuum gaps exist between the plasma and the inner wall, $W_1=R_1$.
The inner wall depicts the perfectly conducting neutron star interior.
The plasma is in contact with the neutron star surface. The outer wall
can be suppressed depending on the outer boundary conditions. For
instance, we expect electromagnetic wave generation, leading to a net
outgoing Poynting flux carrying energy to infinity. Both situations
will be considered.  We generalise the study presented in
\cite{2007A&A...469..843P} by including particle inertia.

\begin{figure*}[htbp]
  \centering 
  \begin{tabular}{cc}
    \includegraphics[scale=0.8]{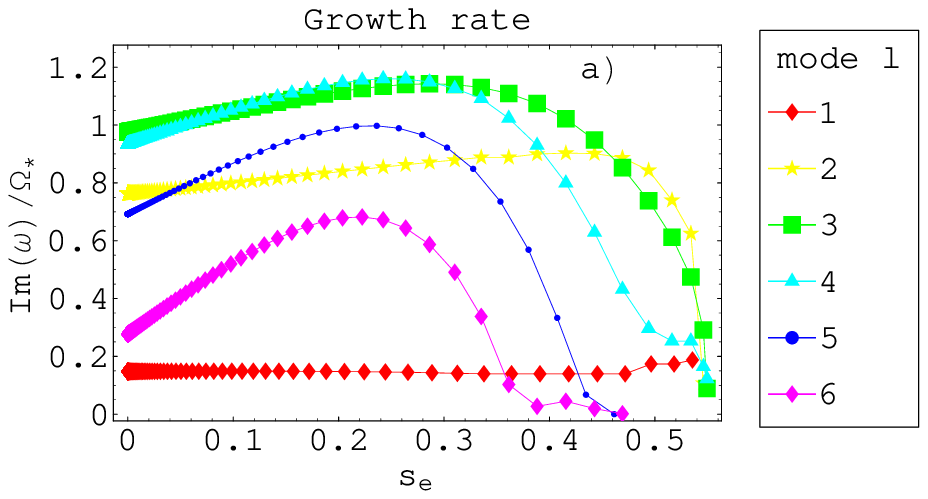} & 
    \includegraphics[scale=0.8]{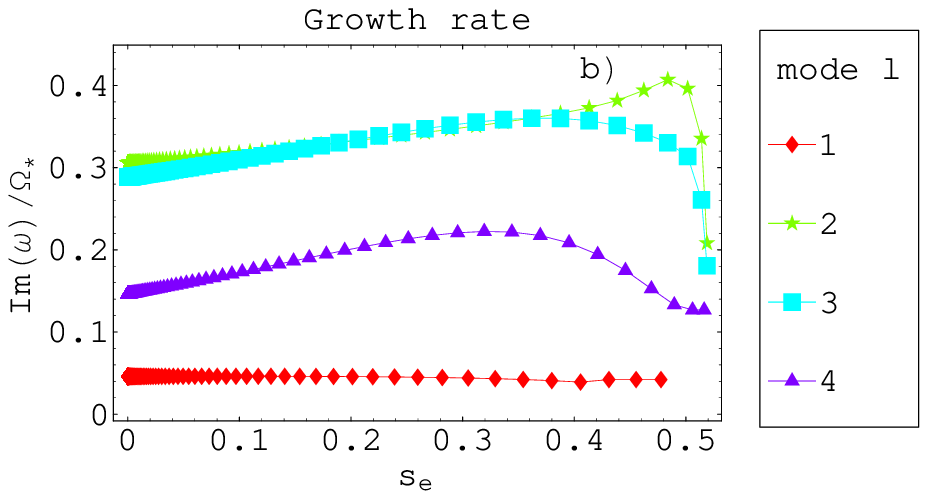}
  \end{tabular}
  \caption{Evolution of the growth rates,~${\rm Im}(\omega)$,
    of the profiles $\Omega_2$, Fig.(a), and $\Omega_3$, Fig.(b), for
    increasing self-field parameter, i.e. transition from the
    diocotron to the magnetron regime. } 
  \label{fig:Elec2}
\end{figure*}

To remain fully self-consistent, we only consider an uniform applied
external magnetic field.  The growth rates for the rotation
curve~$\Omega_2$ for each mode~$l$ are shown in Fig.~\ref{fig:Elec2}a
and those for $\Omega_3$ in Fig.~\ref{fig:Elec2}b.  

For $\Omega_2$, starting from the diocotron regime, $s_{\rm e} \ll 1$,
the instability is first enhanced by the particle inertia effect,
until $s_{\rm e} \approx 0.25$. Then, the growth rates begin to
decrease until they vanish for the maximum allowed self-field
parameter, corresponding to the last existing equilibrium
configuration.

For $\Omega_3$, no increase is observed, the stabilisation effect
proceeds as soon as $s_{\rm e}$ increases.

Finally, we compute the evolution of the spectrum of the magnetron
instability when going to the relativistic regime for a significant
initial self-field parameter, $s_{\rm e} \approx 0.16$. We start with
a non-relativistic rotation profile such that $\beta_2 \ll 1$ and
slowly increase $R_2$ (as well as $W_1, W_2, R_1$ to maintain their
ratio constant) in order to approach the speed of light for the
maximal rotation rate of the plasma column.

\begin{figure*}[htbp]
  \centering
  \begin{tabular}{cc}
    \includegraphics[scale=0.8]{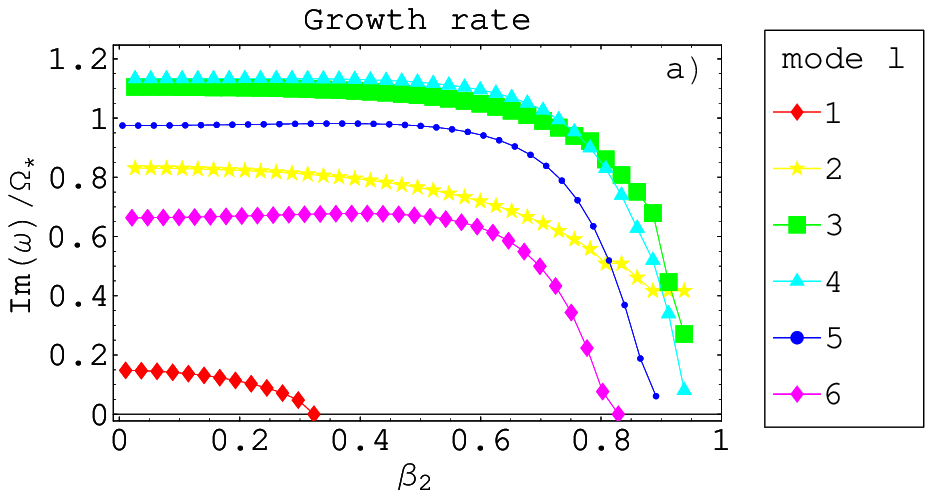} & 
    \includegraphics[scale=0.8]{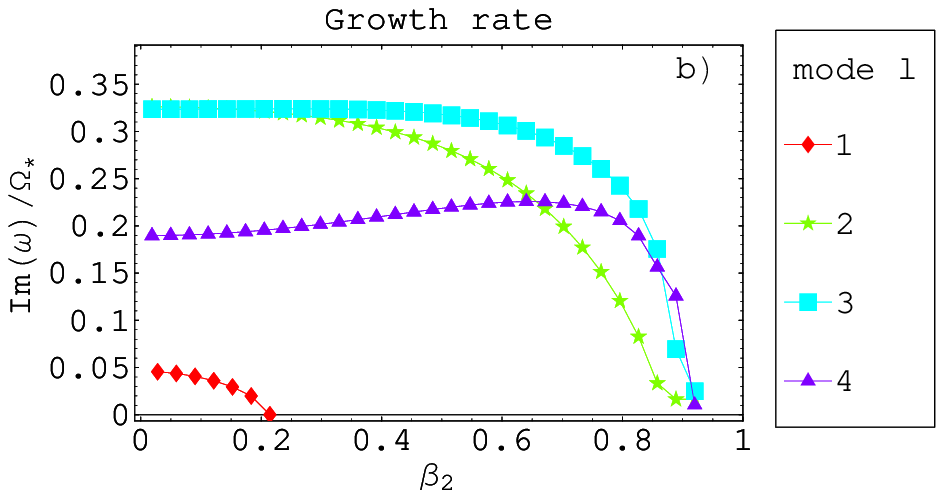}
  \end{tabular}
  \caption{Evolution of the growth rates~${\rm Im}(\omega)$ 
    of the profiles $\Omega_2$, Fig.(a), and $\Omega_3$, Fig.(b), for
    a significant initial self-field parameter, $s_{\rm e} \approx
    0.16$.}
  \label{fig:Elec3}
\end{figure*}

Results for the growth rates of the velocity profiles $\Omega_2$ and
$\Omega_3$ are shown in Fig.~\ref{fig:Elec3}a and~b respectively.

\begin{figure*}[htbp]
  \centering
  \begin{tabular}{cc}
  \includegraphics[scale=0.8]{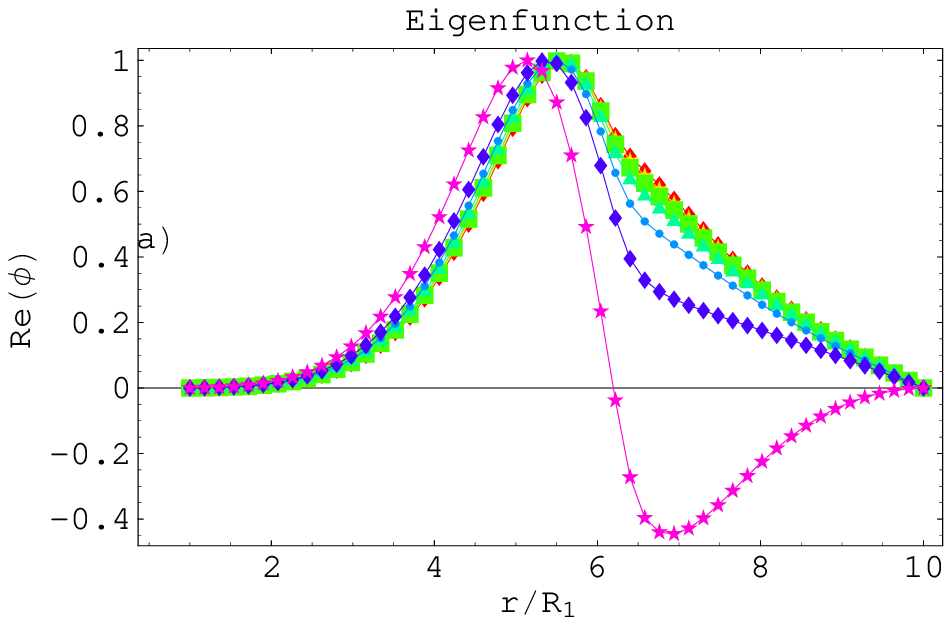} &
  \includegraphics[scale=0.8]{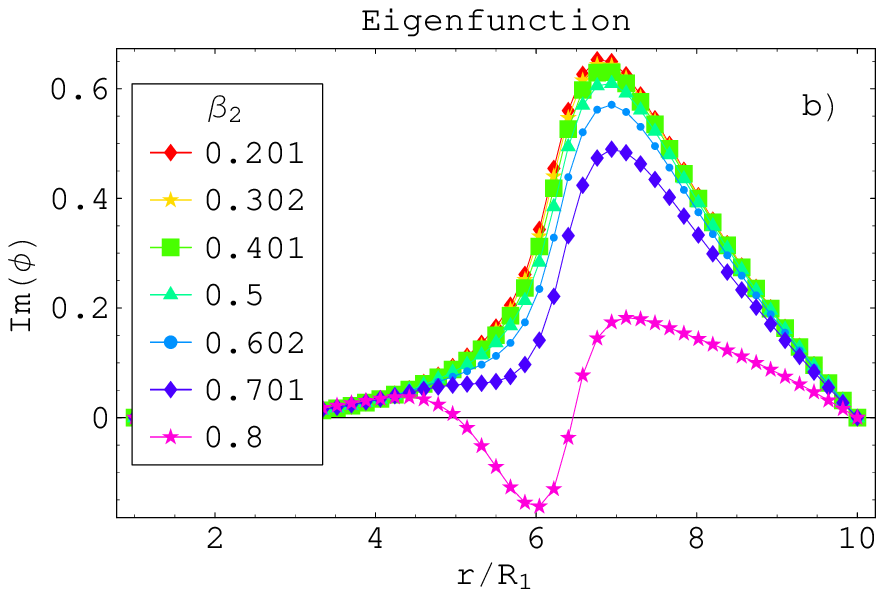}
  \end{tabular}
  \caption{Real and imaginary part of the eigenfunction, 
    respectively Fig.~a and b, for the mode $l=5$ in
    Fig.~\ref{fig:Elec3}a. The real part of the eigenfunction is
    normalised such that the maximum in absolute value is unity. The
    value of the maximal speed~$\beta_2$ for each eigenfunction is
    shown in legend.}
  \label{fig:Elec3_Vec_Prop}
\end{figure*}

For completeness, the evolution of the shape of the eigenfunctions of
the mode $l=5$ corresponding to the eigenvalues in
Fig.~\ref{fig:Elec3}a is shown in Fig.~\ref{fig:Elec3_Vec_Prop}.  When
the stabilisation process starts, the real and imaginary parts of the
eigenfunction oscillate.  Moreover, while these functions remain
positive for small speeds, they change sign whenever the diminishing
of the growth rates become significant. This behavior was already
observed in the study of the influence of the self-field parameter,
Sect.~\ref{subsec:Lab}.

\section{CONCLUSION}
\label{sec:Conclusion}

We developed a numerical code to compute the eigenspectra and
eigenfunctions of the magnetron instability in cylindrical geometry,
including particle inertia, as well as electromagnetic and
relativistic effects, in a fully self-consistent way. It is thus
possible to study the behavior of the plasma in the vicinity of the
light cylinder, i.e. where the particle kinetic energy becomes
comparable to the magnetic field energy density.

Unstable modes are computed for a uniform external applied magnetic
field and arbitrary velocity, density and electric field profiles. The
resulting equilibrium configuration is computed self-consistently,
according to the cold-fluid and Maxwell equations.  Application of the
code to a plasma column as well as to the pulsar electrosphere have
been shown. In both cases, the magnetron regime gives rise to
instabilities that become less and less unstable when the self-field
parameter of the flow, $s_{\rm e}$, approaches its maximum allowed
value, $s_{\rm e} \lesssim s_{\rm e,max}$.  Whereas the growth rates
can be comparable to the rotation period of the neutron star in the
non-relativistic limit, it is found that for special rotation
profiles, the magnetron instability is completely suppressed in the
relativistic regime, as was already the case in the diocotron limit.

However, the magnetron instability is not restricted to the
electrosphere. For instance, it is believed that the pulsed radio
emission emanates from the magnetic poles close to the neutron star
surface. Highly relativistic and dense leptonic flows are accelerated
at these polar caps, forming non-neutral plasma beams able to radiate
coherent electromagnetic waves by several processes like the cyclotron
maser or the free electron laser instability. We will address this
problem as well as the effect of finite temperature in a forthcoming
paper.

\begin{acknowledgements}
  This work was supported by a grant from the G.I.F., the
  German-Israeli Foundation for Scientific Research and Development.
\end{acknowledgements}

\end{document}